\documentclass[12pt]{JHEP3}
\usepackage{amsfonts}
\usepackage{amssymb}
\usepackage{lscape}
\usepackage{cite}
\usepackage{epsfig}
\usepackage{multirow}
\usepackage{longtable}
\usepackage[all]{xy}
\usepackage{amsmath}

\newcommand{\be}{\begin{equation}}
\newcommand{\ee}{\end{equation}}
\newcommand{\ba}{\begin{array}}
\newcommand{\ea}{\end{array}}
\newcommand{\bea}{\begin{eqnarray}}
\newcommand{\eea}{\end{eqnarray}}

\newcommand{\ov}{\overline}
\def\IR{\relax{\rm I\kern-.18em R}}

\def\IP{\relax{\rm I\kern-.18em P}}
\def\inbar{\vrule height1.5ex width.4pt depth0pt}
\def\IC{\relax\,\hbox{$\inbar\kern-.3em{\rm C}$}}

\def\K3{{\bf K3}}

\def\ov{\overline}

\def\n2d{\cN_{V^*}^{\otimes 2}}

 \def\ie{{\it i.e.}~}
 \def\eg{{\it e.g.}~}

\def\IC{\mathbb{C}}

\def\IR{\mathbb{R}}

\def\IP{\mathbb{P}}

\def\cN{{\mathcal N}}

\def\nn{\nonumber}

\title{On closed-string twist-field correlators \\ and their open-string descendants}
\author{
Pascal Anastasopoulos$^{1}$\footnote{pascal@hep.itp.tuwien.ac.at},~
Massimo Bianchi$^{2}$\footnote{Massimo.Bianchi@roma2.infn.it},~
Robert Richter$^{2,3}$\footnote{robert.richter@desy.de   }~\\
$^1$ Technische Univ. Wien Inst. f\"ur Theoretische Physik, A-1040 Vienna, Austria\\
$^2$ Dipartimento di Fisica \& Sezione I.N.F.N., Universit\`a di Roma ``Tor Vergata'', \\
Via della Ricerca Scientica, 00133 Roma, Italy \\
 $^3$ II. Institut f\"ur Theoretische Physik, Hamburg University, Germany\\
}
\date{}

\maketitle
\abstract{
In a recent paper we have proposed the possibility that the lightest
massive string states could be identified with open strings living at
intersections of D-branes forming small angles. In this note, we
reconsider the relevant twist-field correlation functions and perform the
analysis of the sub-dominant physical poles in the various channels. Our derivation is new in that it is based on the algebraic procedure for the construction of open string models
starting from their closed-string `parents' rather than on the stress-tensor method. We also indicate possible generalizations and diverse applications of our approach.}
\preprint{
TUW-11-24\\
ROM2F/2011/16\\
ZMP-HH/11-17}
 \clearpage
\begin{document}

\section{Introduction}

Historically, the first instance of a string amplitude is the Veneziano amplitude, describing the scattering of four `would-be' pions but actually tachyons. Yet, thanks to the powerful constraints of conformal invariance and modular invariance in the perturbative regime, a systematic construction has been successfully achieved for closed-string theories  much earlier than for open-string theories \cite{Green:1987sp, Green:1987mn, Polchinski:1998rq, Polchinski:1998rr, Kiritsis:2007zza}. The systematic construction of theories with open unoriented strings was completed in the early 90's \cite{Bianchi:1990yu, Bianchi:1990tb}\footnote{For a review see \eg \cite{Angelantonj:2002ct}.}. The advent of D-branes \cite{Angelantonj:2002ct} triggered an enormous interest in the subject that led to the construction of exactly solvable models with D-branes and $\Omega$-planes based on tori, orbifolds, free fermions, minimal models, group manifolds and Gepner models \cite{Bianchi:1988ux, Bianchi:1988fr, Bianchi:1989du, Bianchi:1990yu, Bianchi:1990tb, Bianchi:1991rd, Bianchi:1991eu, Angelantonj:1996mw, Angelantonj:1996uy,
Pradisi:1995qy, Recknagel:1997sb, Blumenhagen:2003su, Blumenhagen:2004cg, Dijkstra:2004ym, Anastasopoulos:2006da}. Most of the attention has been devoted to the spectrum \ie to the one-loop partition function.

Much less work has been devoted to the computation of scattering amplitudes in non-trivial settings based on genuinely interacting CFT's \cite{Bianchi:1991rd, Angelantonj:1996mw, Pradisi:1995qy, Recknagel:1997sb, Blumenhagen:2003su, Blumenhagen:2004cg, Dijkstra:2004ym, Anastasopoulos:2006da}. The only prominent exceptions are twist field correlators on tori and orbifolds \cite{Gava:1997jt, David:2000um, David:2000yn, Cvetic:2003ch,Abel:2003vv,Abel:2003yx}. As reviewed in \cite{Light}, twist fields are essential ingredients in the vertex operators for open string states living at brane intersections. However one should keep in mind that while closed-string twists are quantized (rational numbers $k/n$), intersection angles are not, since they depend on both discrete (wrappings) and continuous (moduli) parameter.

Aim of the present note is to derive open-string twist-field correlators using algebraic techniques, devised in \cite{Bianchi:1990yu, Bianchi:1990tb} and applied to minimal models \cite{Bianchi:1991rd} and later on to WZW models \cite{Pradisi:1995qy}, and complementing the `stress tensor method' \cite{Gava:1997jt, David:2000um, David:2000yn, Cvetic:2003ch,Abel:2003vv,Abel:2003yx}. In this approach one has to carefully decompose the `parent' closed-string amplitude in conformal blocks and then take the `square root' in each sector, thus producing chiral blocks evaluated at a real argument that are to be combined with appropriate coefficients. The latter are tightly constrained by `planar duality' \cite{Bianchi:1991rd} and by correct factorization, \ie correct normalization and positivity of the residues of the poles.

We consider correlators of four bosonic twist-fields with one or two independent angles, recalling the closed-string computation and then deriving the open string results. In order to take the relevant square-root, we need to rewrite the closed-string result as a sesqui-linear form in the conformal blocks. We comment on possible ambiguity and arbitrariness in the procedure. As mentioned before, open-string twist-field can be considered `descendants' of closed-string ones only for rational intersection angles. Since rational form a dense set, one can safely extend the result to irrational values (in units of $\pi$) for the angles.

We conclude with preliminary considerations on how to generalize our approach to more sophisticated cases (such as WZW or Gepner models) that are not related in such a simple way as orbifolds to free CFT's.

\section{General strategy \label{sec general strategy}}

In the works of \cite{Gava:1997jt, David:2000um, David:2000yn, Cvetic:2003ch,Abel:2003vv,Abel:2003yx} the authors compute open string correlators that contain multiple bosonic twist fields. In order to do so they extend the world-sheet, whose original domain is the upper half complex plane, via the ``doubling trick" to the full complex plane and employ conformal field theory techniques that are related to the study of bosonic twist fields in the context of closed string theory on orbifolds \cite{Dixon:1986qv,Burwick:1990tu}.

Here we follow a different approach based on the systematic construction of open-string theories from their parent closed-string  theories \cite{Bianchi:1990yu, Bianchi:1990tb}. This algebraic procedure has been mostly applied to the determination of the spectrum coded in the four contributions -- torus, Klein-bottle, annulus and M\"obius-strip -- to the one-loop partition function.
One notable extension of the systematic procedure is the computation of 4-point correlation functions of open strings at tree-level (disk) in minimal models that display the desired factorization properties and `planar duality \cite{Bianchi:1991rd}. Later on this computation has been generalized to $SU(2)$ WZW models \cite{Pradisi:1995qy}.

Quite remarkably both one-loop amplitudes and 4-point correlators at tree level depend on one `invariant', which is complex for oriented closed strings and `real' for open and unoriented strings. At one-loop one has the complex modular parameter $\tau$ for the torus, that becomes purely imaginary ($\tau = i \tau_2$) for Klein-bottle and Annulus and has a fixed real part ($\tau_1= 1/2$) for the M\"obius-strip \cite{Bianchi:1988ux, Bianchi:1988fr, Bianchi:1989du, Bianchi:1990yu, Bianchi:1990tb}. At tree level, one has the complex cross-ratio $z=z_{12}z_{34}/z_{13}z_{24}$ for the sphere and a real cross ratio $x=x_{12}x_{34}/x_{13}x_{24}$ for the upper half plane, topologically equivalent to a disk.

As we will momentarily see, the close similarity between tree-level 4-point correlators and one-loop partition functions is more than a mere analogy. Indeed, twist-field correlators on the sphere can be expressed as (fake) torus partition functions once an appropriate branched-cover of the sphere is considered that trivializes the monodromies.

Before entering the details of the twist-field correlators, let us describe our strategy
\cite{BianchiPHD, StanevMaster, Fuchs:1999fh, Fuchs:2000hn}. Generically a closed string correlator takes the form
\begin{align}\nn
{\cal A}_{closed} &= \langle \Phi_{h_1,\ov h_1}(z_1, \bar{z}_1) \Phi_{h_2,\ov h_2}(z_2, \bar{z}_2)\Phi_{h_3,\ov h_3}(z_3, \bar{z}_3)\Phi_{h_4,\ov h_4}(z_4, \bar{z}_4)\rangle \\ &= \prod_{i<j} |z_{ij}|^{-(h_i + \ov h_i +h_j +\ov h_j)+{\Delta +\ov \Delta \over 3}}\sum_{i,j} c_{ij} {\cal F}_i(z)   \ov {\cal F}_j (\ov z)
\end{align}
where $\Delta= h_1 + h_2 +h_3 +h_4$, $\ov \Delta= \ov h_1 + \ov h_2 + \ov h_3 +\ov h_4$ and  $ {\cal F}_i$ and $ \ov {\cal F}_j$ denote the holomorphic and anti-holomorphic conformal blocks, respectively. The index $i$ runs over the conformal families that appear in the s-channel \ie in the OPE's of both $\Phi_{h_1}\Phi_{h_2}$ and $\Phi_{h_3}\Phi_{h_4}$ exposed in the limit $z\rightarrow 0$. Clearly one can express the same amplitude in the t- or u- channels that give rise to different OPE's and thus different bases of conformal blocks. A defining property of the conformal blocks ${\cal F}_i(z)$ is that, up to an overall factor $z^{h-h_1-h_2}$, they admit an expansion in integer powers of $z$ very much as one-loop characters $\chi_h(w)$ admit an expansion in integer powers of $w=\exp(2\pi i \tau)$, up to an overall factor $w^{h-{c\over 24}}$.

Closely following the systematic construction of one-loop amplitudes, open string 4-point correlators take the form
 \begin{align}
{\cal A}_{open} = \langle \phi_{h_1}(x_1) \phi_{h_2}(x_2) \phi_{h_3}(x_3) \phi_{h_4}(x_4)\rangle = \prod_{i<j} x_{ij}^{-(h_i  +h_j)+{\Delta  \over 3}}\sum_{i} a_i{\cal F}_i(x)
\end{align}
where $x$ is real. The coefficients $a_i$ are tightly constrained by planar duality \cite{Bianchi:1991rd, Pradisi:1995qy} and are adjusted in such a way that they give rise to the correct pole structures. For our purpose some specific limits correspond to the exchange of the identity sector or other untwisted states.

In the simple case of $Z_2$ twist-fields, the above procedure was essentially followed in \cite{Antoniadis:1993jp}. For generic twists, the closed string bosonic correlator with one or two independent angles takes the form
\begin{align}
{\cal A}_{closed} = |K(z)|^2 \, \sum_{\vec{k},\vec{v}} c_{\vec{k}\vec{v}} w(z)^{\frac{\alpha' p^2_L}{4}} \,\,\ov w(\ov z)^{\frac{\alpha' p^2_R}{4}}\,\,,
\label{eq closed string form}
\end{align}
where $w(z)$ is a holomorphic function and $p_L$ and $p_R$ denote the closed string momenta
\begin{align}
p^2_L =\left(\vec{k}+ \frac{\vec{v}}{\alpha'}\right)^2 \qquad \qquad p^2_R = \left(\vec{k}- \frac{\vec{v}}{\alpha'}\right)^2\,\,.
\end{align}
Note that $\vec{k}$ and $\vec{v}$ are two dimensional vectors in the dual lattice $\Lambda^*$ and the lattice $\Lambda$, respectively.
Then the corresponding open string result will be given by
\begin{align}
{\cal A}_{open} = K(x) \, \sum_{p,q} c_{p,q} w(x)^{\alpha' p^2_{open}}
\end{align}
with $p$ and $q$ being integers and $p^2_{open}$ denoting the open string mass. For a D1-brane wrapping a one-cycle on a two torus the mass is given
by  \cite{Kors:2001ku,Angelantonj:2005hs}
\begin{align}
p^2_{open}= \frac{1}{L^2} p^2 +\frac{1}{\alpha'^2}\frac{R^2_1 R^2_2}{L^2} q^2\,\,.
\end{align}
Here $R_1$ and $R_2$ are the radii of the two torus and $L$ denotes the length of the brane the open string is attached to. The first term corresponds to  the  Kaluza-Klein states along the brane while the latter to the winding excitations along the perpendicular direction.

Thus one of the tasks in our approach is to bring the closed string bosonic twist field correlator in the form of  \eqref{eq closed string form}. Such an expression allows one then to make an ansatz for the open string correlator as laid out above.
Given this ansatz one furthermore has to determine the coefficients $c_{p,q}$ by analyzing specific limits that correspond to the exchange of universal states like the gauge bosons (identity sector) or their excitations in the untwisted sector. Positivity of the residues of the poles in all channels puts severe constraints on the $c_{p,q}$ and guarantees `planar duality' \cite{Bianchi:1991rd, Pradisi:1995qy}.

\section{Bosonic twist field correlator with one independent angle}

Generically the closed string correlator is determined via the energy momentum tensor method and given in Lagrangian form. As already mentioned before in order to extract the open string correlator from the closed string one one has to manipulate the closed string result in such a way that it reveals the form \eqref{eq closed string form}. Specifically that implies a Poiss\`on resummation in one of the lattice variables.

\subsection{Closed string correlator}
Using the energy-momentum tensor method the  quantum part of the closed string bosonic twist field correlator with one independent angle has been shown to be \cite{Dixon:1986qv}
\begin{align}
|z_{\infty}|^{2 \theta (1-\theta)}\,\,\langle \sigma_{1-\theta}(0)\, \sigma_{\theta}(z,\ov z) \,
\sigma_{1-\theta}(1) \, \sigma_{\theta} (\infty)\rangle = {\cal C} \,\, \frac{|z(1-z)|^{-2\theta(1-\theta)}}{F(z)\ov F(1-\ov z) +\ov F( \ov z) F(1-z)}\,\,,
\end{align}
where $0< \theta < 1$ is a rational number encoding the twist ($k/n$) and $F(z)$ denotes the hypergeometric function
\begin{align}
F(z)={_2F}_1[\theta,1-\theta,1,z] \,\,.
\end{align}
The action for the classical solutions is given by (here  and in the following we set $\alpha'=2$ for the closed string computations)
\begin{align}
S_{cl}(v_1,v_2)=\frac{\pi}{4 \tau_2 \sin(\pi \theta)} \big[ v_2 \ov v_2 + \tau_1 \left( v_1 \ov v_2 \ov \beta + \ov v_1 v_2 \beta  \right) + |\tau|^2 v_1 \ov v_1\big]\,\,,
\end{align}
where $\beta=-i e^{-i\pi \theta}$ denotes a phase and $\tau(z)$ denotes the modulus of a ``fake torus'' introduced in \cite{Dixon:1986qv}
\begin{align}
\tau(z) =\tau_1+i\tau_2=i\frac{F(1-z)}{F(z)}  \qquad \qquad \qquad \ov \tau(\ov z) =\tau_1-i\tau_2=-i\frac{\ov F(1-\ov z)}{\ov F(\ov z)}\,\,.
\label{eq definition of tau}
\end{align}
Here $v_1$ and $v_2$ denote vectors belonging to some coset of the lattice $\Lambda$.

Combining both results gives
\begin{align}
|z_{\infty}|^{2 \theta (1-\theta)}\,\, \langle \sigma_{1-\theta}(0)\, \sigma_{\theta}(z,\ov z) \,
\sigma_{1-\theta}(1) \, \sigma_{\theta} (\infty)\rangle = {\cal C} \,\, \frac{|z(1-z)|^{-2\theta(1-\theta)}}{\tau_2(z,\ov z) |F(z)|^2} \,\, \sum_{v_1,v_2} e^{-S_{cl}(v_1,v_2)}\,\,,
\end{align}
where we used \eqref{eq definition of tau} to simplify the quantum part. This is the Lagrangian form of the correlator. In the following we will perform a Poiss\`on resummation over the variable $v_2$ to obtain the Hamiltonian form which is much more convenient to extract the corresponding open string result.

After Poiss\`on resumming one ends up with a sum over momenta in the dual lattice $\Lambda^*$, however since the sum over $v_2$ is only over a subset of the lattice $\Lambda^*$ we will substitute $v_2=-2\,\beta^{-1}\,\sin(\pi \theta) \left( f_{23} + q \right)$, where the summation over $q$ is now over the entire lattice $\Lambda$.  Poiss\`on resummation over $q$ yields
\begin{align}
\label{eq hamiltonian form closed 1 angle}
 |z_{\infty}|^{2 \theta (1-\theta)}\,\, \langle \sigma_{1-\theta}(0)\, \sigma_{\theta}(z,\ov z) \,
\sigma_{1-\theta}(1) \, \sigma_{\theta}& (\infty)\rangle = \frac{{\cal C}}{V_{\Lambda}} \,\, \frac{|z(1-z)|^{-2\theta(1-\theta)}}{ |F(z)|^2}\\ \nn
\times &\sum_{k \in \Lambda^*, v_1 \in \Lambda_c} \exp\left[-2 \pi i f_{23} \cdot k \right] \,\, w(z)^{1/2(k+\frac{v}{2})^2}\,  \ov w(\ov z)^{1/2(k-\frac{v}{2})^2},
\end{align}
where $\Lambda_c= (1-\theta)(f_{21}+\Lambda)$ denotes the coset over which $v$ runs, $V_{\Lambda}$ is the volume of the unit cell of the lattice $\Lambda$, and  $w(z)$ is given by
\begin{align}
w(z)=\exp\left[ \frac{i\pi \tau(z)}{\sin(\pi \theta)}\right]\,\,.
\end{align}
Note that the result \eqref{eq hamiltonian form closed 1 angle} takes the form \eqref{eq closed string form} that allows us to extract the open string result.

\subsection{Open string correlator}

Before we extract the open string correlator from the closed string one, let us briefly discuss such a correlator in the context of intersecting D-branes on a compactified six-torus $T^6=T^2 \times T^2 \times T^2$. In this work we assume that all branes go through the origin of the respective two-tori, thus all Wilson lines vanish. Moreover, we further simplify the setup by assuming that both branes $a$ and $b$ wrap such one-cycles on the respective two-tori that they intersect exactly once. Our results can be easily generalized to the more generic case of arbitrary number of intersections and non-vanishing Wilson-lines. We refer the interested reader to \cite{Cremades:2003qj,Cvetic:2003ch,Abel:2003yx, Abel:2003vv,Cremades:2004wa} for more details.

\begin{figure}[h]
\begin{center}
\includegraphics[height=5cm]{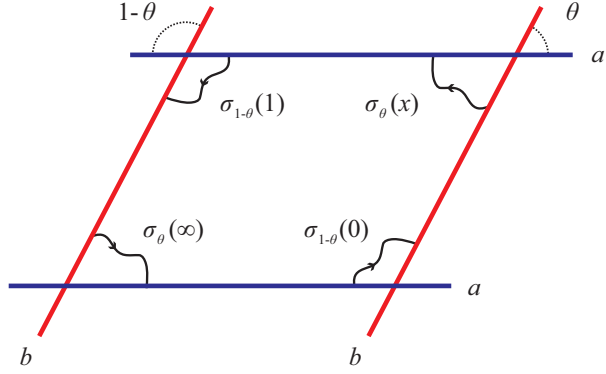}
\end{center}
\caption{Intersection of two D-brane stacks, $a$ and $b$, in one complex dimension.}
\label{fig twistoneangle}
\end{figure}

Figure \ref{fig twistoneangle}  depicts the intersection of two D-branes labeled by $a$ and $b$ on one of the three two-tori. At the intersection there exists a massless fermion going from brane $b$ to brane $a$, whose vertex operator contains the bosonic twist field $\sigma_{\theta}$, where $\theta$ denotes the intersection angle in one two torus.
At the cost of being pedantic, contrary to the twist in the `parent' closed string, $\theta$ is not quantized in the open-string case.

At the very same intersection there will be its antiparticle, a string going from brane $a$ to brane $b$, whose vertex operator contains a anti-twist field $\sigma_{1-\theta}$ \footnote{For a detailed discussion on vertex operators of massless states for arbitrary intersection angles, see \cite{Cvetic:2006iz,Bertolini:2005qh}, for a generalization to massive states see \cite{Light} and for a discussion on instantonic states at the intersection of  D-instanton and D-brane at arbitrary angles, see \cite{Cvetic:2009mt}. For an analysis of vertex operators for massive states in heterotic string theories, see \cite{Bianchi:2010es}.}.

Even though in this work we only consider the twist field correlator and not the amplitude of four matter fields that would contain additional conformal fields, it is worthwile to notice that the non fixed twist field operator position $x$ should remain in the interval $[0,1]$ since otherwise the Chan-Paton factors make the whole amplitude vanishing.

Finally, before we  perform the steps laid out in section \ref{sec general strategy} to derive the open string correlator let us discuss our expectations for the two limits $x \rightarrow 0$ and  $x \rightarrow 1$. In both limits we expect the exchange of untwisted string states, just as one observes in the closed string correlator. We will use gauge-boson exchange to normalize the correlator.

The result \eqref{eq hamiltonian form closed 1 angle} is of the type that allows us to extract the open string correlator. Since the open string result should only depend on real parameters we  expect $w(x)$ to take the form
\begin{align}
w(x)=\exp\left[- \frac{\pi t(x)}{\sin(\pi \theta)}\right]
\label{eq def w(x) one angle}
\end{align}
with $t(x)$ given by  \eqref{eq definition of tau}
\begin{align}
t(x) = \frac{1}{2 i} \left( \tau(x) - \ov \tau(x )\right)= \frac{F(1-x)}{F(x)}\,\,.
\end{align}
In the last step we used the fact that $\ov F(x) = F(x)$ for $0\leq x\leq 1 $.
Concerning the quantum part one expects only the holomorphic part to survive. Finally the mass of the open strings in the untwisted sector is given by \cite{Kors:2001ku,Angelantonj:2005hs}
\begin{align}
M^2_{open}(p,q)=  \frac{1}{L^2_a} p^2+ \frac{1}{\alpha'^2}\frac{R^2_1 \, R^2_2}{L^2_a} q^2\,\,,
\label{eq mass KK and winding}
\end{align}
where $R_1, R_2$ are the radii of the two-torus and $L_a$ is the length of the D-brane
\begin{align}
L_a = \sqrt{n^2_a R^2_1+ m^2_a R^2_2 }
\end{align}
with $n_a$, $m_a$ being the wrapping numbers of the brane $a$ on the two torus.

Thus the open string result takes the form
\begin{align}
x^{ \theta (1-\theta)}_{\infty} \langle \sigma_{1-\theta}(0)\, \sigma_{\theta}(x) \,
\sigma_{1-\theta}(1) \, \sigma_{\theta} (\infty)\rangle = {\cal C}_{open} \frac{[x(1-x)]^{-\theta(1-\theta)}}{F(x)} \sum_{p,q} w(x)^{ \left(  \frac{\alpha'}{L^2_a}  p^2 + \frac{1}{\alpha'}\frac{R^2_1 \, R^2_2}{L^2_a} q^2 \right)}
\label{eq suggested result open one angle}
\end{align}
with $w(x)$ given in \eqref{eq def w(x) one angle}.

This ansatz has to satisfy various non-trivial constraints, that generalize `planar duality', namely in the limit $x\rightarrow 0$ as well as for $x \rightarrow 1$ we expect an exchange of untwisted states. This condition also allows us to fix the constant ${\cal C}_{open}$.

Let us start with the limit $x \rightarrow 0$. In that limit $t(x)$ behaves
\begin{align}
\lim_{x\rightarrow 0} t(x)  =\frac{1}{\pi} \sin(\pi \theta)\, \left( -\ln(x) +\ln(\delta)\right)
\end{align}
with
\begin{align}
\ln(\delta) = 2 \psi(1) -\psi(\theta)- \psi(1-\theta)\,\,.
\end{align}
Plugging in that limit one obtains
\begin{align}
 {\cal C}_{open} x^{-\theta(1-\theta)} \sum_{p,q} \left(\frac{x}{\delta}\right)^{ \left(  \frac{\alpha'}{L^2_a}  p^2 + \frac{1}{\alpha'}\frac{R^2_1 \, R^2_2}{L^2_a} q^2 \right)}
 \label{eq result one angle a}
\end{align}

Let us turn to the $x \rightarrow 1$ limit which also exhibits the exchange of untwisted states, corresponding to strings starting and ending on the brane $b$ rather than on brane $a$. In order to determine this limit we perform a Poiss\`on resummation in the variables $p$ and $q$ and obtain
\begin{align}
{\cal C}_{open} \frac{[x(1-x)]^{-\theta(1-\theta)}}{F(x)} \, \frac{\sin(\pi \theta)}{t(x)} \frac{L^2_a}{R_1 \, R_2} \, \sum_{\tilde{p} \in \Lambda_{p}, \tilde{q} \in  {\Lambda}^*_{q}} \exp\left[-\frac{\pi \sin(\pi \theta)}{t(x)} \left( \frac{L^2_a}{\alpha'} \tilde{p}^2 +\frac{ \alpha'\, L^2_a}{R^2_1\, R^2_2} \tilde{q\,}^2 \right) \right]
\end{align}
With the identification (recall that brane $a$ and brane $b$  intersect exactly once on the torus)
\begin{align}
\sin(\pi \theta) = \frac{R_1 R_2}{L_a L_b}
\end{align}
this further simplifies to
\begin{align}
{\cal C}_{open}  \frac{[x(1-x)]^{-\theta(1-\theta)}}{F(1-x)} \, \frac{L_a}{L_b} \,\, \sum_{\tilde{p} \in \Lambda_{p}, \tilde{q} \in  \Lambda^*_{q}} \exp\left[-\frac{\pi }{\sin(\pi \theta) t(x)} \left( \frac{R^2_1 R^2_2}{ \alpha' \, L^2_b} \tilde{p}^2 +\frac{\alpha'}{L^2_b} \tilde{q}^2 \right) \right]\,\,.
\end{align}
Now performing the limit $x \rightarrow 1$ yields
\begin{align}
 {\cal C}_{open} \frac{L_a}{L_b} (1-x)^{-\theta(1-\theta)} \sum_{\tilde{p},\tilde{q}} \left(\frac{1-x}{\delta}\right)^{ \left(  \frac{\alpha'}{L^2_b} \,  \tilde{q}^2+ \frac{R^2_1 \, R^2_2}{\alpha' L^2_b} \, \tilde{p}^2\right)}\,\,.
\label{eq result one angle b}
\end{align}
Comparing the limits \eqref{eq result one angle a} and \eqref{eq result one angle b} it is natural to assume that the normalization constant is given by  ${\cal C}_{open}= \frac{\sqrt{\alpha'}}{L_a}$.

Let us now compare the derived open string expression to the ones in \cite{Gava:1997jt, David:2000um, David:2000yn, Cvetic:2003ch,Abel:2003vv,Abel:2003yx} derived via the energy momentum tensor method after extending the world-sheet to the whole complex plane. The first thing to notice is that \eqref{eq suggested result open one angle} is given in the Hamiltonian form while the correlator in \cite{Gava:1997jt, David:2000um, David:2000yn, Cvetic:2003ch,Abel:2003vv,Abel:2003yx} is of Lagrangian type. Thus we have to Poiss\`on resum \eqref{eq suggested result open one angle} over the variable $p$ which gives
\begin{align}
& \sqrt{\sin(\pi \theta)}\frac{[x(1-x)]^{-\theta(1-\theta)}}{\sqrt{F(x) F(1-x)}} \\ & \hspace{1cm}\times \sum_{\tilde{p} \in \Lambda_{p}, \tilde{q} \in  {\Lambda}^*_{q}} \exp\left[-\frac{\pi}{\alpha'} \sin(\pi \theta) F(x) F(1-x) \left(\left( \frac{\tilde{p} L_a}{F(1-x)}\right)^2+
\left( \frac{q L_b}{F(x)}\right)^2\right) \right] \,\,.\nn
\end{align}
This is indeed the result obtained via employing conformal field theory techniques. We conclude that in this case there is no ambiguity in the choice of the coefficients of the linear combination of conformal blocks.

Before we turn to the bosonic twist correlator with two independent angles let us determine the sub-dominant poles for the correlator \eqref{eq suggested result open one angle} in the limit $x \rightarrow 0$. This requires the knowledge of the behaviour of $F(x)$ for small $x$, which is (see appendix \ref{app hypergeometric})
\begin{align}
\lim_{x\rightarrow 0} F(x) = \frac{1}{\Gamma(\theta) \, \Gamma(1-\nu)} \sum^{\infty}_{n=0} \frac{\Gamma(\theta+n) \, \Gamma(1-\nu+n)}{\Gamma(n)}  \frac{x^n}{n !}\,\,.
\end{align}
Given that the correlator \eqref{eq suggested result open one angle} behaves in the limit $x \rightarrow 0$ as
\begin{align}
\frac{\sqrt{\alpha'}}{L_a} x^{-\theta(1-\theta)} \left( 1 - \theta(1-\theta) x -\frac{1}{2} \theta (1-\theta)  ((\theta^2 -  \theta + 2) x^2+ ... \right)
 \sum_{p,q} \left(\frac{x}{\delta}\right)^{ \left(  \frac{\alpha'}{L^2_a}  p^2 + \frac{1}{\alpha'}\frac{R^2_1 \, R^2_2}{L^2_a} q^2 \right)}
\end{align}
suggests the following OPE
\begin{align}
\sigma_{\theta}(z) \, \sigma_{1-\theta} (w) & \sim \sqrt{\frac{\sqrt{\alpha'}}{L_a}} (z-w)^{-\theta(1-\theta)} \,\, \mathbf{1} \\ \nn
& \hspace{15mm} + \sqrt{\frac{\sqrt{\alpha'}}{L_a}} \sqrt{\theta(1-\theta)} (z-w)^{-\theta(1-\theta)+1} \left( e^{i\alpha} \partial X +   e^{-i\alpha} \partial \ov X \right) +...
\end{align}
Here the phase $\alpha$ indicates the arbitrariness in the definition of the conformal fields $\sigma$ as well as $\partial X$ and $\ov \partial X$, and can be eliminated by an appropriate redefinition of the latter. One can easily extend this OPE to higher order poles which involve then conformal fields with larger conformal dimension.

\section{Bosonic twist field correlator with two independent angles}

Now we turn to the bosonic twist correlator with two independent angles. We follow the same procedure as above for the correlator with just one independent angle.  We will manipulate the closed string result to obtain an expression of type \eqref{eq closed string form} and extract from that the open string result.

\subsection{Closed string correlator}

Let us display the complete result, containing the quantum as well as the classical part, for the four twist field correlator with two independent angles determined in 
\cite{Burwick:1990tu}\footnote{See also \cite{Stieberger:1992bj,Erler:1992gt}.} (again we set $\alpha'=2$)
\begin{align}
\label{eq correlator 2 independent angles}
& |z_{\infty}|^{2 \nu (1-\nu)}\,\,\langle \sigma_{1-\theta}(0)\, \sigma_{\theta}(z,\ov z) \,
\sigma_{1-\nu}(1) \, \sigma_{\nu} (\infty)\rangle \\ & \hspace{2cm} =|z|^{-2\theta(1-\theta)}
\left(1-z\right)^{-\nu(1-\theta)}\, \left(1-\ov z\right)^{-\theta(1-\nu)}
I^{-1}( z, \bar{z}) \,\,\sum_{v_1, v_2}e^{-S_{cl}} \nn
\end{align}
with $I(z, \bar{z})$ given by
\begin{align*}
I(z, \bar{z})= \frac{1}{2\pi} \big[B_1(\theta,\nu) \,G_2( z)
\ov H_1(1-\ov z)+ B_2(\theta,\nu) \, \ov G_1(\ov z) {H_2}(1- z)\big]\,\,,
\end{align*}
where
\begin{align*}
\begin{gathered}
B_1(\theta,\nu)=\frac{\Gamma(\theta)\,\Gamma(1-\nu)}{\Gamma(1+\theta-\nu)}\qquad
B_2(\theta,\nu)=\frac{\Gamma(\nu)\,\Gamma(1-\theta)}{\Gamma(1+\nu-\theta)}\\
\\
G_1(z)= {_2F}_1[\theta,1-\nu,1;z]\qquad
G_2(z)= {_2F}_1[1-\theta,\nu,1;z]\\
\\
H_1(z)= {_2F}_1[\theta,1-\nu,1+\theta-\nu;z]\qquad
 H_2(z)={_2F}_1[1-\theta,\nu,1-\theta+\nu;z]\,\,.
 \end{gathered}
\end{align*}
and $S_{cl}$ is
\begin{align}
S_{cl}=V_{11} v_1 \ov v_1 +V_{12} v_1 \ov v_2 +V^*_{21} v_2 \ov v_1 +V_{22} v_2 \ov v_2 \,\,.
\end{align}
Here the $V_{ij}$ are given by
\begin{align} \nn
V_{11} &= \frac{1}{4} \left(\frac{\sin(\pi \theta)}{\pi}\right)^2 |I(z,\ov z)|^{-2} \\ & \nn
 \hspace{1cm}\left( \,\, B_2 \big|H_2(1-z)\big|^2\Big[ B_1 G_1(z) \ov H_1(1-\ov z)+  B_1 \ov G_1(\ov z)  H_1 (1- z) \right.
\\& \left.\nn
  \hspace{5cm}
  + \pi \left( \cot(\pi \nu) - \cot(\pi \theta) \right)\left| G_1(z)\right|^2\Big]
\right. \\ & \left.   \nn
 \hspace{1cm} +B_1 \big|H_1(1-z)\big|^2\Big[ B_2 G_2(z) \ov H_2(1-\ov z)+  B_2\ov G_2(\ov z)  H_2 (1- z)
 \right. \\ &\left.
   \hspace{5cm}
  + \pi \left( \cot(\pi \theta) - \cot(\pi \nu) \right)\left| G_2(z)\right|^2\Big] \right)
\end{align}
\begin{align} \nn
V_{12} &= \frac{e^{i\pi \theta}}{4} \frac{\sin(\pi \theta)}{\pi} |I(z,\ov z)|^{-2} \\ & \nn
 \hspace{1cm}\left( \,\, B_2 G_2 (z) \ov H_2(1-\ov z)\Big[ B_1 G_1(z) \ov H_1(1-\ov z)+  B_1 \ov G_1(\ov z)  H_1 (1- z) \right.
\\& \left.\nn
  \hspace{5cm}
  + \pi \left( \cot(\pi \nu) - \cot(\pi \theta) \right)\left| G_1(z)\right|^2\Big]
\right. \\ & \left.   \nn
 \hspace{1cm} -B_1 \ov G_1(\ov z ) \ov H_1(1-z)\big|^2\Big[ B_2 G_2(z) \ov H_2(1-\ov z)+  B_2\ov G_2(\ov z)  H_2 (1- z)
 \right. \\ &\left.
   \hspace{5cm}
  + \pi \left( \cot(\pi \theta) - \cot(\pi \nu) \right)\left| G_2(z)\right|^2\Big] \right)
\end{align}
\begin{align} \nn
V_{22} = \frac{1}{4} |I(z,\ov z)|^{-2} \,\, &\left( \,\, G_1(z) G_2 (z) \Big[ B_1 \ov G_2(\ov z) \ov H_1(1-\ov z)+  B_2 \ov G_1(\ov z)  H_2 (1-\ov z) \right.
\\& \left.
 \hspace{0.1cm} + \ov G_1( \ov z) \ov G_2 (\ov z) \Big[ B_1 G_2( z) H_1(1- z)+  B_2  G_1(z)  H_2 (1-z)\right)\,\,.
\end{align}
In order to extract the open-string correlator from the closed-string one let us perform a Poiss\`on resummation, just as we did for the four bosonic twist correlator with just one independent angle. The substitution $v_2 = 2 i e^{-i\pi \theta} \sin(\pi \theta) \left( f_{23} + q \right)$ yields to a summation over the lattice $\Lambda$
\begin{align}
\sum_{q, v_1 \in \Lambda} e^{-S_{cl}}   = \sum_{q, v_1 \in \Lambda} &\exp\Big[ -V_{11} |v_1|^2 + 2 i\, \sin(\pi \theta) e^{i\pi \theta}  V_{12}\,\, v_1  \left( \ov f_{23} + \ov q \right)  \\ \nn
&   -2 i\, \sin(\pi \theta) e^{-i\pi \theta}  V^*_{12}\,\, \ov v_1  \left(  f_{23} + q \right)  - 4 \sin^2(\pi \theta) V_22 \,\, \left(  f_{23} + q \right) \left( \ov f_{23} + \ov q \right) \Big]\,\,,
\end{align}
which allows us to perform Poiss\`on resummation. One obtains (note that here we omit the quantum part)
\begin{align*}
\frac{\pi}{4 \sin^2(\pi \theta) V_{22}}
\,\,\sum_{k \in \Lambda^*, v \in \Lambda}  & \exp\Big[ -\frac{\pi^2}{4 \sin^2(\pi \theta) V_{22} }
|\vec{k}|^2  - 2 \pi i \vec{f}_{23} \cdot \vec{k} + \pi \frac{\left( e^{-i \pi \theta} V_{12} -e^{i \pi \theta}  {V_{12}}^*\right)}{2 \sin( \pi \theta)\,\, V_{22} } \, \vec{k} \cdot \vec{v}\\
& + i \pi \frac{\left( e^{-i \pi \theta} V_{12} +e^{i \pi \theta} {V_{12}}^*\right)}{2 \sin( \pi \theta)
\,\, V_{22} }  \vec{p}^{\,T}
\left(\begin{array}{cc}
0&1\\
-1& 0
\end{array}
\right) \vec{v} + \left(\frac{V_{12}  {V_{12}}^*}{V_{22}} -V_{11} \right) |\vec{v}|^2
 \Big]\,\,.
\end{align*}

Some statements are in order. First of all note that $v_1 \rightarrow v$. Moreover, in the last line the $\vec{k}$ and $\vec{v}$ are two-dimensional vectors rather than complex numbers. In addition one can verify the following, very useful relations
\begin{align}
\left(\frac{V_{12}  {V_{12}}^*}{V_{22}} -V_{11} \right)  = -\frac{\pi^2}{16 \sin^2(\pi \theta) V_{22} }
\end{align}
\begin{align}
 \frac{\left( e^{-i \pi \theta} V_{12} +e^{i \pi \theta} {V_{12}}^*\right)}{2 \sin( \pi \theta)
\,\, V_{22} } = \frac{1}{4} \big( \cot(\pi \nu) - \cot(\pi \theta) \big)\,\,.
\end{align}
Finally, it is convenient to define the following quantities
\begin{align}
\tau_1(z,\ov z) = i \, \frac{1}{2 \, V_{22}} \left({V_{12}}^* - V_{12} \right)
\end{align}
and
\begin{align}
\tau_{2} (z, \ov z) =\frac{\pi}{4 \, \sin(\pi \theta)} \,\, \frac{1}{V_{22}} \,\,.
\end{align}
They combine to the holomorphic ({\it a priori} far from obvious) quantity
\begin{align}
\tau(z) =\tau_1(z,\ov z) + i \tau_{2} (z, \ov z)  = i \frac{\sin(\pi \theta)}{2\, \pi} \left( \frac{B_1 H_1(1-z)}{G_1(z)} + \frac{B_2 H_2(1-z)}{G_2(z)}\right) \,\,,
\end{align}
which for the special case of equal angles, $\theta =\nu$, takes the form \eqref{eq definition of tau}. With these identifications we can write the Poiss\`on resummed expression as (here we still omit the quantum part)
\begin{align}
\frac{\pi}{4 \sin^2(\pi \theta) V_{22}}
\,\,\sum_{k \in \Lambda^*, v \in \Lambda}  e^{i\pi \vec{k}^T B \vec{v} }  e^{-2 \pi i \vec{f}_{23} \cdot \vec{k}}\,\, w(z)^{1/2(k+v/2)^2} \,\, \ov w(\ov z)^{1/2(k-v/2)^2}
\label{eq hamiltonian without quantum}
\end{align}
with
\begin{align}
B= \frac{1}{4} \big( \cot(\pi \nu) - \cot(\pi \theta) \big) \,\, \left(\begin{array}{cc}
0&1\\
-1& 0
\end{array} \right)
\end{align}
and
\begin{align}
w(z)=\exp\left[{i\frac{\pi \tau(z)}{\sin(\pi \theta)}} \right]\,\,.
\end{align}
Adding the quantum part (see eq. \eqref{eq correlator 2 independent angles}) to \eqref{eq hamiltonian without quantum} one obtains
\begin{align}
|z_{\infty}|^{2 \nu (1-\nu)}\,\,\langle \sigma_{1-\theta}(0)\, \sigma_{\theta}(z,\ov z) \,
\sigma_{1-\nu}(1) \, \sigma_{\nu} (\infty)\rangle  =&
|z|^{-2\theta(1-\theta)}
|1-z|^{-\theta(1-\nu)}\, \frac{1}{|G_1(z)|^2}  \\ & \hspace{-2cm} \times
\,\,\sum_{k \in \Lambda^*, v \in \Lambda}  e^{i\pi \vec{k}^T B \vec{v} }  e^{-2 \pi i \vec{f}_{23} \cdot \vec{k}}\,\, w(z)^{\frac{(k+v/2)^2}{2}} \,\, \ov w(\ov z)^{\frac{(k-v/2)^2}{2}}\,\,. \nn
\label{eq poisson resum closed result}
\end{align}
This is exactly  the same as the complicated expression \eqref{eq correlator 2 independent angles}. Apart from the obviously much simpler form it is also of the type \eqref{eq closed string form} that allows us to extract the open string result. Before we turn to this derivation we analyze the two interesting limits $z \rightarrow 0$ and $z \rightarrow 1$.

In the limit $z \rightarrow 0$ we expect the exchange  of untwisted states. In that limit $w(z)$ behaves as
\begin{align}
\lim_{z \rightarrow 0} w(z) = \frac{z}{\delta}
\end{align}
with $\ln \delta= 2\psi(1) - \psi(\theta) -\psi(1-\theta) -\psi(\nu) - \psi(1-\nu) $. Here we used various relations and limits displayed in appendix \ref{app hypergeometric}.

This gives the expected result which looks similar to the limits $z \rightarrow 0$ and $z \rightarrow 1$ for the closed string correlator with just one independent twist/angle \cite{Dixon:1986qv}
\begin{align}
|z|^{-2\theta(1-\theta)}
\,\,\sum_{k \in \Lambda^*, v \in \Lambda}  e^{i\pi \vec{k}^T B \vec{v} }  \frac{e^{-2 \pi i \vec{f}_{23} \cdot \vec{k}}}{\delta^{h+\ov h}}\,\,z^h\,\ov z^{\ov h}
\end{align}
with the conformal weights $h$ and $\ov h$ given by
\begin{align}
h=\frac{1}{2}\left(k+\frac{v}{2}\right)^2\qquad \qquad  \qquad \ov h=\frac{1}{2}\left(k-\frac{v}{2}\right)^2\,\,.
\end{align}

In contrast in the limit $z \rightarrow 1$ we expect the exchange of twisted states. Thus the closed string twist field correlator OPE's can be derived from taking the limit
$z \rightarrow 1$ of \eqref{eq correlator 2 independent angles} where we ignore the classical part of the correlator. For $z \rightarrow 1$ the quantum part of \eqref{eq correlator 2 independent angles} behaves as
\begin{align}
\label{eq expansion nu>theta}
2 \pi \,|1-z|^{-2 \theta(1-\nu)} &\frac{\Gamma(1-\theta) \, \Gamma( \nu) \, \Gamma(\nu-\theta)}{\Gamma(\theta) \, \Gamma( 1-\nu) \, \Gamma(1+\theta-\nu)}\\  & \times  \left[ 1-\frac{\Gamma^2(1-\theta) \, \Gamma^2(\nu) \, \Gamma(1+\theta-\nu) \, \Gamma(\theta -\nu) }{\Gamma^2(\theta) \, \Gamma^2(1-\nu) \, \Gamma(1-\theta +\nu) \, \Gamma(\nu-\theta) } |1-z|^{2\nu-2\theta} + ...\right] \nn
\end{align}
for $\nu >\theta$ and\footnote{The angle $\alpha$ appearing in the bosonic twist field is in the open interval $(0,1)$. Thus for $\nu >\theta$ the resulting angle on the right-hand side is not the naively expected angle $1-\theta+\nu$, but rather $\nu-\theta$.}
\begin{align}
\label{eq expansion nu<theta}
2 \pi \,|1-z|^{-2 \nu(1-\theta)} &\frac{\Gamma(\theta) \, \Gamma( 1-\nu) \, \Gamma(1-\theta+\nu)}{\Gamma(1-\theta) \, \Gamma( \nu) \, \Gamma(\theta-\nu)}\\  & \times  \left[ 1-\frac{\Gamma^2(\theta) \, \Gamma^2(1-\nu) \, \Gamma(\nu- \theta) \, \Gamma(1-\theta +\nu) }{\Gamma^2(1-\theta) \, \Gamma^2(\nu) \, \Gamma(\theta -\nu) \, \Gamma(1+\theta-\nu) } |1-z|^{2\theta-2 \nu} + ...\right] \nn
\end{align}
for $\theta > \nu$. Using  the properties displayed in appendix \ref{app hypergeometric} it is straightforward to show that hypergeometric functions behave as
\begin{align} \nn
&\lim_{z\rightarrow 1} G_1(z) = \frac{\Gamma(\nu-\theta)}{\Gamma(1-\theta) \, \Gamma(\nu)} +(1-z)^{\nu-\theta} \frac{\Gamma(\theta-\nu)}{\Gamma(\theta) \, \Gamma(1-\nu)}\\
&\lim_{z\rightarrow 1} G_2(z) = \frac{\Gamma(\theta-\nu)}{\Gamma(\theta) \, \Gamma(1-\nu)} +(1-z)^{\theta-\nu} \frac{\Gamma(\nu-\theta)}{\Gamma(1-\theta) \, \Gamma(\nu)}\\
&\lim_{z\rightarrow 1} H_1(1-z)  = \lim_{z\rightarrow 1} H_2(1-z) =1\nn\,\,.
\end{align}
which eventually gives rise to the limits \eqref{eq expansion nu>theta} and \eqref{eq expansion nu<theta}.

Thus the OPE of two bosonic twist fields is given by
\begin{align}
\sigma_{\theta} (z, \ov z) \, \sigma_{1-\nu} (w, \ov w) \sim\,\, & C^-_{\sigma}\,  \,|z-w|^{-2 \theta(1-\nu)}\,\sigma_{1+\theta-\nu}\,  (w, \ov w) \\
& \hspace{10mm}+  C^-_{\widetilde{\tau}} \,\, |z-w|^{-2 \theta(2-\nu)+2 \nu} \widetilde{\tau}_{1+\theta-\nu}\, (w , \ov w)\,  \nn
\end{align}
for $\nu > \theta$, and takes the form
\begin{align}
\sigma_{\theta} (z, \ov z) \, \sigma_{1-\nu} (w, \ov w) \sim\,\, & C^+_{\sigma}\,  \,|z-w|^{-2 \nu(1-\theta)}\,\sigma_{\theta-\nu}\,  (w, \ov w) \\
& \hspace{10mm}+  C^+_{{\tau}} \,\, |z-w|^{-2 \nu(2-\theta)+2 \theta} {\tau}_{\theta-\nu}\, (w , \ov w)\,  \nn
\end{align}
for $\theta > \nu$. Here the coefficients are
\begin{align}
\label{eq OPE closed yukawa1}
C^-_{\sigma} &= \sqrt{2 \pi \,\,\frac{\Gamma(1-\theta)\,\Gamma(\nu)\, \Gamma(1+\theta-\nu)}{\Gamma(\theta)\,\Gamma(1-\nu)\, \Gamma(\nu-\theta)}}\\
C^-_{\tau} &=\sqrt{ 2 \pi \,\,\frac{\Gamma^3(1-\theta)\,\Gamma^3(\nu)\, \Gamma^2(1+\theta-\nu)\, \Gamma(\theta-\nu)}{\Gamma^3(\theta)\,\Gamma^3(1-\nu)\, \Gamma^2(\nu-\theta)\, \, \Gamma(1-\theta+\nu)}}\\
\label{eq OPE closed yukawa2}
C^+_{\sigma} &= \sqrt{2 \pi\,\, \frac{\Gamma(\theta)\,\Gamma(1-\nu)\, \Gamma(1-\theta+\nu)}{\Gamma(1-\theta)\,\Gamma(\nu)\, \Gamma(\theta-\nu)}}\\
C^+_{\widetilde{\tau}} &=\sqrt{ 2 \pi \,\,\frac{\Gamma^3(\theta)\,\Gamma^3(1-\nu)\, \Gamma^2(1-\theta+\nu)\, \Gamma(\nu-\theta)}{\Gamma^3(1-\theta)\,\Gamma^3(\nu)\, \Gamma^2(\theta-\nu)\, \, \Gamma(1+\theta-\nu)}}\,\,,
\end{align}
where \eqref{eq OPE closed yukawa1} and \eqref{eq OPE closed yukawa2} are related to the quantum part of the physical Yukawa couplings determined in
\cite{Burwick:1990tu,Stieberger:1992bj}. The  weights of the respective conformal fields are given by
\begin{align}
h_{\sigma_{\alpha}} &= \frac{1}{2} \alpha \left(1-\alpha \right) \qquad \qquad \qquad \hspace{11mm}\ov h_{\sigma_{\alpha}} = \frac{1}{2} \alpha \left(1-\alpha \right)\\
h_{\widetilde{\tau}_{\alpha}} &= \frac{1}{2} \left(1-\alpha \right)\left(2+\alpha \right) \qquad \qquad \qquad \ov h_{\widetilde{\tau}_{\alpha}} = \frac{1}{2} \left(1-\alpha \right)\left(2+\alpha \right)\\
h_{\tau_{\alpha}} &= \frac{1}{2} \alpha \left(3-\alpha \right) \qquad \qquad \qquad \hspace{11mm}\ov h_{\tau_{\alpha}} = \frac{1}{2} \alpha \left(3-\alpha \right)\,\,.
\end{align}
Let us make a few remarks regarding this result. So far we have not distinguished between the bosonic twist field $\sigma^+_{\alpha}$ and the bosonic anti-twist field $\sigma^{-}_{\alpha}$. As shown in the appendix \ref{app twists} the anti-twist field $\sigma^{-}_{\alpha}$ can be identified with  $\sigma^+_{1-\alpha}$. Thus for the case at hand, namely the OPE $\sigma_{1-\theta}(z, \ov z) \, \sigma_{\nu}(w, \ov w) $, it is not surprising to obtain two different scenarios, depending on which angle is larger. One should interpret that OPE as an OPE between a twist field and an anti-twist field. Depending on the choice angles these two fields couple to a bosonic twist field or an bosonic anti-twist field. Similar arguments apply to the the sub-dominant couplings to the excited twist fields $\tau_{\alpha}(z, \ov z)$ and $\widetilde{\tau}_{\alpha}(z, \ov z )$. That suggests that $\tau_{\alpha}(z, \ov z)$
should be interpreted as $\tau^+_{\alpha}(z, \ov z)$, while $\widetilde{\tau}_{\alpha}(z, \ov z)$
should be interpreted as excited anti-twist field $\tau^-_{\beta}(z, \ov z)$, with $\beta= 1-\alpha$. This is also exposes the analogy between the two fields, since they have now the same conformal dimension $h_{\tau^+_{\alpha}}=\ov h_{\tau^+_{\alpha}}=h_{\tau^-_{\alpha}}=\ov h_{\tau^-_{\alpha}}=\frac{1}{2} \alpha(3-\alpha)$.

One can further extend the OPE's \eqref{eq OPE closed yukawa1} and \eqref{eq OPE closed yukawa2} by looking at higher order terms in the expansions  \eqref{eq expansion nu>theta} and \eqref{eq expansion nu<theta}, respectively. One finds that the two bosonic twist fields couple to  doubly excited twist fields as well. This can be generalized to the statement that they couple to N-times excited twist fields for any $N$. As we will see momentarily this is in contrast to the open string, where one observes an even grading, namely the bosonic twist fields couple only to  N-times excited twist fields where $N$ is even.

\subsection{Open string correlator \label{sec open two angles}}

\begin{figure}[h]
\begin{center}
\includegraphics[height=5cm]{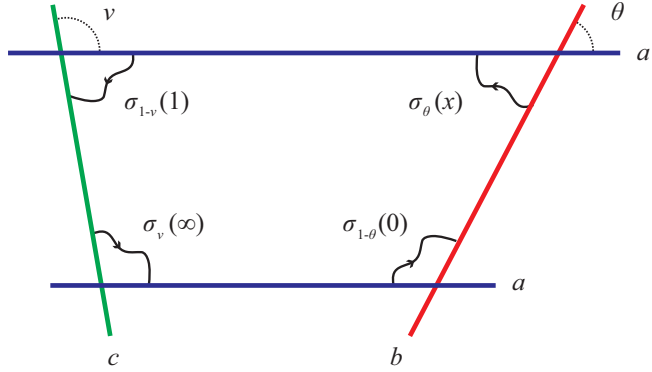}
\end{center}
\caption{Intersection of three D-brane stacks, $a$, $b$ and $c$, in one complex dimension.}\label{fig twisttwoangles}
\end{figure}

In contrast to the previously discussed case the setup here consists of three D-branes denoted by $a$, $b$ and $c$. Again we assume a simplified setup in which all D-branes go through the origin and all three D-branes intersect each other once. Figure \ref{fig twisttwoangles} depicts the above described setup for one of the three two-tori. The intersection angle between $a$ and $b$ is given by $\theta$ while between  $a$ and $c$  it is $\nu$. As for the correlator with just one independent angle only for the interval $0 \leq x \leq 1$ the result is non-vanishing\footnote{This statement is slightly modified in un-oriented theories in which one of the D-branes is the orientifold image of one of the other two. Such a situation has been discussed in \cite{Cvetic:2006iz}.}.

In contrast to the correlator with just one independent angle here only in the limit $x \rightarrow 0$ we expect the exchanges of untwisted states. This limit allows us to normalize the correlator. On the other hand in the limit $x \rightarrow 1$ one observes the exchange of twisted states. As for the closed string correlator this limit allows us then to determine the proper operator product expansion of two bosonic twist fields even beyond the dominant pole.

Following the same steps as for the four bosonic twist field correlator with just one independent angle we obtain for the correlator
 \begin{align} \nn
 & x^{\nu (1-\nu)}_{\infty}\,\,  \langle \sigma_{1-\theta}(0)\, \sigma_{\theta}(x) \,
\sigma_{1-\nu}(1) \, \sigma_{\nu} (\infty)\rangle = \\ & \hspace{2cm} {\cal C}_{open} \frac{x^{-\theta(1-\theta)} (1-x)^{-\theta(1-\nu)}}{G_1(x)} \sum_{p,q} w(x)^{ \left(  \frac{\alpha'}{L^2_a}  p^2 + \frac{1}{\alpha'}\frac{R^2_1 \, R^2_2}{L^2_a} q^2 \right)}
\label{eq suggested result open two angles}
 \end{align}
with $w(x)$ given by
\begin{align}
w(x)=\exp\left[- \frac{\pi t(x)}{\sin(\pi \theta)}\right]
\end{align}
where $t(x)$ is
\begin{align}
t(x) = \frac{\sin(\pi \theta)}{2 \pi} \left( \frac{B_1 H_1(1-x)}{G_1(x)} + \frac{B_2 H_2(1-x)}{G_2(x) }
 \right) \,\,.
 \end{align}
Here we use the fact that  $G_i (x) = \ov G_i (x) $ and $H_i (x) = \ov H_i (x)$ for $0\leq x\leq 1$. Note that, as expected, \eqref{eq suggested result open two angles} reduces to \eqref{eq suggested result open one angle} for $\theta= \nu$ .

Let us determine the constant ${\cal C}_{open}$ which we of course expect to be $\sqrt{\alpha'}/L_a$. Recall that we expect the following OPE
\begin{align}
\sigma_{1-\theta}(z) \, \sigma_{\theta} (w) \sim (z-w)^{-\theta(1-\theta)} \,\,\frac{\sqrt{\alpha'}}{L_a} \,\,\mathbf{1}\,\,,
\label{eq expected result}
\end{align}
where the normalization factor $\sqrt{\alpha'}/L_a$ is crucial for consistency of the four bosonic twist field correlator with just one independent angle. In order to determine the  constant ${\cal C}_{open}$ we analyze the limit $x \rightarrow 0 $, which corresponds to the exchange of untwisted states. In that limit $t(x)$ behaves like
\begin{align}
t(x) \approx \frac{\sin (\pi\theta)}{ \pi}\left( -\ln(x) + \ln(\delta)\right)
\end{align}
with $\ln (\delta)$ given by
\begin{align}
\ln(\delta) = 2 \psi(1) -\frac{1}{2} \left(\psi(\theta) + \psi(1-\theta) + \psi(\nu) + \psi(1-\nu) \right)\,\,.
\end{align}
That leads to the following expression for \eqref{eq suggested result open two angles} in the limit $x \rightarrow 0$
\begin{align}
{\cal C}_{open}\, x^{-\theta(1-\theta)} \sum_{p,q} \left(\frac{x}{\delta}\right)^{\left(  p^2 \frac{\alpha'}{L^2_a} + q^2 \frac{R^2_1 \, R^2_2}{\alpha'\, L^2_a}\right)}
\label{eq lim x -> 0 two angles}
\end{align}
Thus comparing \eqref{eq lim x -> 0 two angles} for $p=q=0$ with \eqref{eq expected result}  we indeed get for the normalization constant
${\cal C}_{open} =\sqrt{\alpha'}/L_a$.

Eventually we are interested in the limit $x \rightarrow 1$ that corresponds to the exchange of twisted states. In order to take the limit it is convenient to perform a Poiss\`on resummation over the variable $p$, such that both variables we sum over live in the lattice $\Lambda$. We obtain (taking already into account the normalization constant)
\begin{align}
\frac{x^{-\theta(1-\theta)} (1-x)^{-\theta(1-\nu)}}{G_1(x)}  \sqrt{\frac{\sin(\pi \theta)}{t(x)}}
\sum_{\widetilde{p},q}
\exp\left[-\pi \frac{\sin(\pi \theta)} {t(x)} \, \frac{L^2_a}{\alpha'} \, \widetilde{p}^2- \pi \frac{t(x) }{\sin(\pi \theta)}  \frac{ R^2_1\, R^2_2}{\alpha' \,  L^2_a}  \, q^2\right]\,\,.
\label{eq Poisson resummed result}
\end{align}
A comparison with the results obtained in \cite{Cvetic:2003ch,Abel:2003vv,Abel:2003yx} shows that both methods, the one using the doubling trick to extend the world-sheet to the complex plane and then using conformal field theory techniques and the other one in which one extracts the open string correlator directly from the closed string correlator, give the exactly same result. Once again we conclude that there are no ambiguity in the choice of the constants $c_{p,q}$ in this case either.

For the limit $x \rightarrow 1$ one has to distinguish two different cases
\begin{align} \label{eq limits t(x) theta>nu}
\lim_{x\rightarrow 1}  t (x) &= \frac{\sin(\pi(\theta-\nu) )}{2 \sin(\pi \nu)} \qquad  \qquad \text{for} \qquad \theta > \nu\\
\label{eq limits t(x) theta<nu}
\lim_{x\rightarrow 1} t (x) &= \frac{\sin(\pi(\nu-\theta) )}{2 \sin(\pi \nu)} \qquad  \qquad \text{for} \qquad \theta < \nu\,\,.
\end{align}
This can be easily derived from
\begin{align} \nn
&\lim_{x\rightarrow 1} G_1(x) = \frac{\Gamma(\nu-\theta)}{\Gamma(1-\theta) \, \Gamma(\nu)} +(1-x)^{\nu-\theta} \frac{\Gamma(\theta-\nu)}{\Gamma(\theta) \, \Gamma(1-\nu)}\\ \label{eq limits of hypergeometric}
&\lim_{x\rightarrow 1} G_2(x) = \frac{\Gamma(\theta-\nu)}{\Gamma(\theta) \, \Gamma(1-\nu)} +(1-x)^{\theta-\nu} \frac{\Gamma(\nu-\theta)}{\Gamma(1-\theta) \, \Gamma(\nu)}\\
&\lim_{x\rightarrow 1} H_1(1-x)  = \lim_{x\rightarrow 1} H_2(1-x) =1\nn\,\,.
\end{align}
Thus for $\nu>\theta$ we get \eqref{eq suggested result open two angles}
\begin{align}
(1-x)^{-\theta(1-\nu)} &\frac{\Gamma(1-\theta)}{\Gamma(\nu)}{\Gamma(\nu -\theta)} \sqrt{\frac{2 \sin(\pi \theta) \, \sin(\pi \nu)}{\sin(\pi (\nu-\theta))}} \\
&\times \sum_{\widetilde{p},q}  \exp\left[-2\pi \frac{\sin(\pi \theta) \, \sin(\pi \nu) }{\sin(\pi(\theta-\nu)} \, \frac{ L^2_a}{\alpha'}\, \widetilde{p}^2- \frac{\pi}{2} \frac{\sin(\pi (\theta-\nu))}{\sin(\pi \theta)\,\sin(\pi \nu) }  \, \,\frac{ R^2_1\, R^2_2}{ \alpha' \,L^2_a} q^2\right]\,\,. \nn
\end{align}

The angles can be expressed in the following matter (recall that all the branes intersect exactly once)
\begin{align}
\sin(\pi \theta) =  \frac{R_1 R_2}{ L_a L_b} \hspace{1cm} \sin(\pi \nu) =  \frac{R_1 R_2}{ L_c L_a} \hspace{1cm} \sin(\pi (\nu-\theta)) =  \frac{R_1 R_2}{ L_b L_c}\,\,.
\label{eq sinus definitions}
\end{align}
Given \eqref{eq sinus definitions} one obtains
\begin{align}
\sqrt{2 \pi}     (1-x)^{-\theta(1-\nu)} & \sqrt{       \frac{\Gamma(1-\theta)\Gamma(\nu)\Gamma(1+\theta-\nu)}
{\Gamma(\theta)\Gamma(1-\nu)\Gamma(\nu-\theta)} } \sum_{\widetilde{p},q}  \exp\left[-\frac{2\pi}{\alpha'} R_1 R_2 \tilde{p}^2- \frac{\pi}{2 \alpha'} R_1 R_2 q^2\right]
\end{align}
which, with the redefinition of the summation variables (note that with this definition  $r_1$ and $r_2$ are integers)
\begin{align}
r_1 = 2 \tilde{p} +q \qquad \qquad r_2 = 2 \tilde{p}- q
\end{align}
leads to
 \begin{align}
\sqrt{2 \pi}     (1-x)^{-\theta(1-\nu)} & \sqrt{       \frac{\Gamma(1-\theta)\Gamma(\nu)\Gamma(1+\theta-\nu)}
{\Gamma(\theta)\Gamma(1-\nu)\Gamma(\nu-\theta)} } \sum_{r_1, r_2}  \exp\left[-\frac{\pi}{4 \alpha'} R_1 R_2 \left( r^2_1 +r^2_2 \right)\right]\,\,.
\label{eq Yukawa final theta<nu}
\end{align}
Analogously one obtains for $\theta> \nu$
 \begin{align}
\sqrt{2 \pi}     (1-x)^{-\nu(1-\theta)} & \sqrt{       \frac{\Gamma(\theta)\Gamma(1-\nu)\Gamma(1-\theta+\nu)}
{\Gamma(1-\theta)\Gamma(\nu)\Gamma(\theta-\nu)} } \sum_{r_1, r_2}  \exp\left[-\frac{\pi}{4 \alpha'} R_1 R_2 \left( r^2_1 +r^2_2 \right)\right]\,\,.
\label{eq Yukawa final theta>nu}
\end{align}
Again $r_1$ and $r_2$ are positive integers. Note that the classical part is exactly twice the world-sheet instanton contribution to the Yukawa coupling arising from a two-torus determined in \cite{Cremades:2003qj, Abel:2003vv}. Moreover, \eqref{eq Yukawa final theta<nu} and \eqref{eq Yukawa final theta>nu}, quantum and classical part, agree with the results of \cite{Cremades:2003qj,Cvetic:2003ch, Abel:2003vv} where the authors investigated Yukawa couplings in intersecting brane worlds.

As we saw for the closed string correlator, to extract the OPE between two bosonic twist fields it is sufficient to look at the quantum part of the correlator. For large volumes, namely in the limit $R_1, R_2 \rightarrow \infty$ the only contribution comes from $(\widetilde{p}, q) =(0,0)$ and the whole correlator reduces to the quantum part given by\footnote{Here we used also the identities displayed in eq. \ref{eq sinus definitions}.}
\begin{align}
\frac{x^{-\theta(1-\theta)} (1-x)^{-\theta(1-\nu)}}{G_1(x)}  \sqrt{\frac{\sin(\pi \theta)}{t(x)}}
\end{align}
which is exactly the quantum part computed in \cite{Cvetic:2003ch,Abel:2003yx} using the same conformal field theory techniques as for the closed string after extending the world-sheet to the whole complex plane via the doubling trick.

We already showed above that the dominant pole gives the expected Yukawa couplings. Now we would like to take a look at the sub-dominant terms. Again we will start with $\nu>\theta$. Using \eqref{eq limits t(x) theta<nu} and \eqref{eq limits of hypergeometric} one gets
\begin{align}
\label{eq expansion theta<nu}
(1-x)^{-\theta(1-\nu)} & \sqrt{2 \pi} \sqrt{       \frac{\Gamma(1-\theta)\Gamma(\nu)\Gamma(1+\theta-\nu)}
{\Gamma(\theta)\Gamma(1-\nu)\Gamma(\nu-\theta)} } \\ & \left( 1-\frac{\Gamma^2(1-\theta) \Gamma^2(\nu) \Gamma(\theta-\nu) \Gamma(1+\theta-\nu)}{\Gamma^2(\theta) \Gamma^2(1-\nu) \Gamma(\nu-\theta) \Gamma(1-\theta+\nu)} (1-x)^{2(\nu-\theta)} + ... \right)\,\,.\nn
\end{align}
This implies that in the OPE between the two bosonic twist fields there is no coupling to the excited twist field $\tau$ but the first sub-dominant pole rather indicates a coupling to the doubly excited twist field $\rho$. Let us display the OPE explicitly (keep in mind one has to take the square root out of the different coefficients in front of the poles)
\begin{align}
\sigma_{\theta}(w) \sigma_{1-\nu} (z) \sim & \left( 2\pi \frac{\Gamma(1-\theta)\Gamma(\nu)\Gamma(1+\theta-\nu)}
{\Gamma(\theta)\Gamma(1-\nu)\Gamma(\nu-\theta)} \right)^{\frac{1}{4}} (z-w)^{-\theta(1-\nu)}  \sigma_{1+\theta-\nu} (z)   \\
& \hspace{-2cm} + \left( 2\pi \frac{\Gamma^5(1-\theta) \Gamma^5(\nu) \Gamma^2(\theta-\nu) \Gamma^3(1+\theta-\nu)}{\Gamma^5(\theta) \Gamma^5(1-\nu) \Gamma^3(\nu-\theta) \Gamma^2(1-\theta+\nu)} \right)^{\frac{1}{4}} (z-w)^{-\theta(3-\nu)+2 \nu}\widetilde{\rho}_{1+\theta-\nu}(z) + ...\nn
\end{align}
Analogously one obtains for $\theta>\nu$ in the limit $x \rightarrow 1$
\begin{align}
\label{eq expansion theta>nu}
   (1-x)^{-\nu(1-\theta)} & \sqrt{2 \pi}   \sqrt{       \frac{\Gamma(\theta)\Gamma(1-\nu)\Gamma(1-\theta+\nu)}
{\Gamma(1-\theta)\Gamma(\nu)\Gamma(\theta-\nu)} }\\ & \left( 1-\frac{\Gamma^2(\theta) \Gamma^2(1-\nu) \Gamma(\nu-\theta) \Gamma(1-\theta+\nu)}{\Gamma^2(1-\theta) \Gamma^2(\nu) \Gamma(\theta-\nu) \Gamma(1+\theta-\nu)} (1-x)^{2(\nu-\theta)} + ... \right)\nn
\end{align}
which leads to the OPE
\begin{align}
\sigma_{\theta}(w) \sigma_{1-\nu} (z) \sim & \left( 2\pi \frac{\Gamma(\theta)\Gamma(1-\nu)\Gamma(1-\theta+\nu)}
{\Gamma(1-\theta)\Gamma(\nu)\Gamma(\theta-\nu)} \right)^{\frac{1}{4}} (z-w)^{-\nu(1-\theta)}  \sigma_{\theta-\nu} (z)   \\
& \hspace{-2.3cm} + \left( 2\pi \frac{\Gamma^5(\theta) \Gamma^5(1-\nu) \Gamma^2(\nu-\theta) \Gamma^3(1-\theta+\nu)}{\Gamma^5(1-\theta) \Gamma^5(\nu) \Gamma^3(\theta-\nu) \Gamma^2(1+\theta-\nu)} \right)^{\frac{1}{4}} (z-w)^{-\nu(3-\theta)+2 \theta} \,\,\rho_{\theta-\nu}(z) + ...\nn\,\,.
\end{align}
Note that in contrast to the closed string, two bosonic open twist fields $\sigma_{\alpha}(w)$ and  $\sigma_{\beta}(z)$ do not couple to the excited twist field $\tau_{\alpha+\beta}$. Their first sub-dominant pole in the OPE indicates the coupling to the doubly excited twist field $\widetilde{\rho}$ and $\rho$ respectively. This can be generalized by looking at higher order terms in the expansions \eqref{eq expansion theta<nu} and \eqref{eq expansion theta>nu}
to the statement that the two bosonic twist fields do only couple to $N$-times excited twist fields, where $N$ is an even number.

The conformal dimensions of the fields $\sigma_{\alpha}$ and $\widetilde{\rho}_{\alpha}$  and ${\rho}_{\alpha}$ are
\begin{align}
h_{\sigma_{\alpha}}= \frac{1}{2} \alpha\left(1-\alpha \right) \qquad  \hspace{4mm} h_{\widetilde{\rho}_{\alpha}}= \frac{1}{2} \left(1-\alpha\right)\left(4+\alpha \right) \qquad   \hspace{4mm} h_{\rho_{\alpha}}= \frac{1}{2} \alpha\left(5-\alpha \right) \,\,.
\end{align}
Analogously to the closed string result $\widetilde{\rho}_{\alpha}$ can be interpret as a doubly-excited anti-twist field $\rho^-_{\beta}$ with $\beta=1-\alpha$. On the other hand $\rho_{\alpha}$ is the doubly-excited twist field $\rho^-_{\alpha}$. As for the closed string this exhibits the analogy of these two conformal fields due to their same conformal dimension $h_{\rho^+_{\alpha}}=h_{\rho^-_{\alpha}}=\frac{1}{2} \alpha \left(5-\alpha \right)$ \footnote{For a more detailed discussion on this issue see \cite{Light} and appendix \ref{app twists}.}.

\section{Summary}

In this note we revisited the correlator of four open string bosonic twist-fields with one and two independent angles, respectively. In contrast to previous works we extracted the open string correlator directly from the closed string result. This way we provide a non-trivial check for the method advocated in \cite{Gava:1997jt,David:2000um,Cvetic:2003ch,Abel:2003vv} where the authors extend  the world-sheet, whose original domain is the upper half complex plane, via the ``doubling trick'' to the full complex plane and employ conformal field theory techniques.

Given the four-point correlators, we investigated various limits that allowed us to extract the OPE's between two bosonic twist fields beyond the dominant terms. Interestingly one finds, in contrast to the closed string, that two bosonic twist fields do not couple to an excited twist field, but rather to a doubly excited twist field. This can be generalized to
the statement that two open string bosonic twist fields couple only to $N$-times excited twist fields with $N$ being an even positive integer.

Finally, we found an interesting identification between bosonic twist field and anti-twist field even for higher excited twist fields. This allows to compute
the twist field correlator for just one combination of ``twist'' and ``anti-twist'' fields. Any other combination can be determined by the appropriate identifications.

Before concluding we would like to sketch how to generalize our discussion to genuinely interacting albeit rational CFT's. While in the case of twist-fields the number of block was actually infinite, in a RCFT the number of blocks is finite. Conformal blocks can be determined by means of null vectors and the resulting Ward identities. Explicit expressions are available for WZW models and for various minimal models. For RCFT the number of D-branes is at most equal to the number of characters. Given a parent closed-string model, specified by the choice of the one-loop modular invariant combination of characters, there may be many open string `descendants'. By the same token, given a closed-string 4-point correlator we expect different open-string correlators, at least as many as open string descendants. In some simple cases, as minimal models (\eg Ising model), only the simplest open-string descendant with real Chan-Paton charges was considered \cite{Bianchi:1991rd} to give 4-point amplitudes compatible with planar duality. Later on, based on the analysis of $SU(2)$ WZW models \cite{Pradisi:1995qy} it was noticed that complex Chan-Paton factors were also allowed.
It would be interesting to explore the problem of computing open-string correlator based on the knowledge of their parent closed-string correlators in WZW or even better in Gepner models. In the latter case only limited knowledge of chiral blocks is available at present, however. Given their relation to solvable compactifications on CY manifolds this is a rather sad state of affairs.

\section*{Acknowledgements}

We acknowledge M. Cveti{\v c}, F. Fucito, E. Kiritsis, G. Leontaris, J. F. Morales Morera, I. Pesando, G. Pradisi, B. Schellekens, O. Schlotterer, M. Schmidt-Sommerfeld, Ya. Stanev, P. Teresi and T. Weigand for interesting discussions and correspondence. P.~A. is supported by FWF P22000.  P.~A. and R.~R. are grateful to the organizers of the school and workshop of ITN ``UNILHC" PITN-GA-2009-237920 for hospitality during parts of this work. The work of R.~R. was partly supported by the German Science Foundation (DFG) under the Collaborative Research Center (SFB) 676 “Particles, Strings and the Early Universe”.
The work of M.~B. was partially supported by the ERC Advanced Grant n.226455 Superfields, by the Italian MIUR-PRIN contract 20075ATT78, by the NATO grant PST.CLG.978785. M.~B. and P.~A. would like to thank NORDITA, Stockholm for hospitality during the completion of this project.

\newpage
\appendix
\section{Properties of hypergeometric functions
\label{app hypergeometric}}
In this appendix we display various properties of hypergeometric functions that we will use throughout the paper.

The hypergeometric function is given by
\begin{align}
{_2F}_1[\theta,1-\theta,1,z] = \frac{1}{\Gamma(\theta) \, \Gamma(1-\nu)} \sum^{\infty}_{n=0} \frac{\Gamma(\theta+n) \, \Gamma(1-\nu+n)}{\Gamma(n)}  \frac{z^n}{n !}\,\,.
\end{align}
where the series is only convergent for $|z|\leq 1$. Below we display some relations of the hypergeometric functions, starting with
\begin{align}
{_2F}_1[a,b,c,z]=(1-z)^{c-a-b}{_2F}_1c-a,c-b,c,z]\,\,.
\end{align}
For $a+b-c \neq m$, where $m \in \mathbf{Z}$
\begin{align}
{_2F}_1[a,b,c,z]&=\frac{\Gamma(c)\Gamma(c-a-b)}{\Gamma(c-a)\Gamma(c-b)} {_2F}_1[a,b,a+b-c+1,1-z] \\ &\hspace{1cm}
(1-z)^{c-a-b} \, \frac{\Gamma(c)\Gamma(a+b-c}{\Gamma(a)\Gamma(b)} {_2F}_1[c-a,c-b,c-a-b+1,1-z] \,\,.\nn
\end{align}
For $c=a+b$ one obtains
\begin{align}
{_2F}_1[a,b,a+b,z]&=\frac{\Gamma(a+b)}{\Gamma(a)\Gamma(b)}  \sum^{\infty}_{n=0}   \frac{(a)_n (b)_n}{(n!)^2}   \\
& \hspace{1cm }\times \left[ 2\psi(n+1) -\psi(a+n)-\psi(b+n)-\ln(1-z)\right](1-z)^n\,\,, \nn
\end{align}
where $\psi(z)$ is the Digamma function $\psi(z)= \frac{d \ln \Gamma(z)}{dz}$ and $(a)_n$ denotes the Pochhammer's symbol $(a)_n= \frac{\Gamma(a+n)}{\Gamma(a)}$.

\section{The twist and anti-twist fields $\sigma^+_{\theta}$ and $\sigma^-_{\theta}$
\label{app twists} }
In this appendix we discuss the OPE's of bosonic twist fields with the conformal fields $\partial Z$ and $\partial \ov Z$. Let us start with the closed string.
The OPE's of the bosonic twist fields take the form \cite{Dixon:1986qv}  (here we only display the relevant OPE's for our purpose)
\begin{align}
&\partial Z (z) \, \ov \partial \ov Z (\ov z)  \sigma^+_{\theta} (w, \ov w) \sim |z-w|^{2\theta-2} \tau^+_{\theta}(z,\ov z)     \\
&\partial \ov Z (z) \, \ov \partial  Z (\ov z)  \sigma^+_{\theta} (w, \ov w) \sim |z-w|^{-2\theta} \widetilde{\tau}^+_{\theta}(z,\ov z)   \\
&\partial Z (z) \, \ov \partial \ov Z (\ov z)  \sigma^-_{\theta} (w, \ov w) \sim |z-w|^{-2\theta} \tau^-_{\theta}(z,\ov z)      \\
&\partial \ov Z (z) \, \ov \partial  Z (\ov z)  \sigma^-_{\theta} (w, \ov w) \sim |z-w|^{-2\theta-2} \widetilde{\tau}^-_{\theta}(z,\ov z)   \,\,.
\end{align}
One notices right away that this suggests that the bosonic twist field $\sigma^-_{\theta}$ has the same OPE as $\sigma^+_{1-\theta}$, and one often identifies these two. Something very analogous happens also with $\tau^-_{\theta}$ that can be identified with $\tau^+_{1-\theta}$.
While one needs further OPE's to justify that statement note that $\tau^-_{\theta}$ and $\tau^+_{1-\theta}$ have the same conformal dimension $h_{\tau^-_{\theta}}=\ov h_{\tau^-_{\theta}}=h_{\tau^+_{1-\theta}}=\ov h_{\tau^+_{1-\theta}}=\frac{1}{2}(1-\theta)(2+\theta)$, which is a necessary condition for this statement to be true.

Analogously for the open string one has
\begin{align}
&\partial X(z)  \sigma^+_{\theta} (w) \sim (z-w)^{-1+\theta} \tau^+_{\theta}(z)     \\
&\partial \ov X (z)  \sigma^+_{\theta} (w) \sim (z-w)^{-\theta} \widetilde{\tau}^+_{\theta}(z)   \\
&\partial X (z) \,   \sigma^-_{\theta} (w) \sim (z-w)^{-\theta} \tau^-_{\theta}(z)      \\
&\partial \ov X (z) \,  \sigma^-_{\theta} (w) \sim (z-w)^{-1+\theta} \widetilde{\tau}^-_{\theta}(z)   \,\,.
\end{align}
Again the anti-twist field $\sigma^-_{\theta}$ can be identified with the twist field $\sigma^+_{1-\theta}$, with the substitution $\theta \rightarrow 1-\theta$. Also the excited  anti-twist field $\tau^-_{\theta}$ can be identified with an excited twist field $\tau^+_{1-\theta}$ with the same angle replacement. To properly justify that we need additional OPE's but the claim withstands the non-trivial check that $\tau^-_{\theta}$  and $\tau^+_{1-\theta}$ have the same conformal dimension $h_{\tau^-_{\theta}}=h_{\tau^+_{1-\theta}}=\frac{1}{2}(1-\theta) (2+\theta) $. This can be generalized to higher excited bosonic twist fileds such as the doubly excited twist field $\rho$, as has been done at the end of section \ref{sec open two angles}.

\clearpage \nocite{*}

\bibliographystyle{JHEP}

\begin{thebibliography}{10}

\bibitem{Green:1987sp}
M.~B. Green, J.~Schwarz, and E.~Witten, {\it {Superstring Theory. Vol. 1:
  Introduction}}, .

\bibitem{Green:1987mn}
M.~B. Green, J.~Schwarz, and E.~Witten, {\it {Superstring Theory. Vol. 2: Loop
  Amplitudes, Anomalies And Phenomenology}}, .

\bibitem{Polchinski:1998rq}
J.~Polchinski, {\it {String theory. Vol. 1: An introduction to the bosonic
  string}}, .

\bibitem{Polchinski:1998rr}
J.~Polchinski, {\it {String theory. Vol. 2: Superstring theory and beyond}}, .

\bibitem{Kiritsis:2007zza}
E.~Kiritsis, {\it {String theory in a nutshell}}, .

\bibitem{Bianchi:1990yu}
M.~Bianchi and A.~Sagnotti, {\it {On the systematics of open string theories}},
   {\em Phys.Lett.} {\bf B247} (1990) 517--524.

\bibitem{Bianchi:1990tb}
M.~Bianchi and A.~Sagnotti, {\it {Twist symmetry and open string Wilson
  lines}},  {\em Nucl.Phys.} {\bf B361} (1991) 519--538.

\bibitem{Angelantonj:2002ct}
C.~Angelantonj and A.~Sagnotti, {\it {Open strings}},  {\em Phys.Rept.} {\bf
  371} (2002) 1--150, [\href{http://xxx.lanl.gov/abs/hep-th/0204089}{{\tt
  hep-th/0204089}}]. Dedicated to John H. Schwarz on the occasion of his
  sixtieth birthday.

\bibitem{Bianchi:1988ux}
M.~Bianchi and A.~Sagnotti, {\it {The Geometry Of Open String Partition
  Functions}}, .

\bibitem{Bianchi:1988fr}
M.~Bianchi and A.~Sagnotti, {\it {The Partition Function Of The SO(8192)
  Bosonic String}},  {\em Phys.Lett.} {\bf B211} (1988) 407.

\bibitem{Bianchi:1989du}
M.~Bianchi and A.~Sagnotti, {\it {Open Strings And The Relative Modular
  Group}},  {\em Phys.Lett.} {\bf B231} (1989) 389.

\bibitem{Bianchi:1991rd}
M.~Bianchi, G.~Pradisi, and A.~Sagnotti, {\it {Planar duality in the discrete
  series}},  {\em Phys.Lett.} {\bf B273} (1991) 389--398.

\bibitem{Bianchi:1991eu}
M.~Bianchi, G.~Pradisi, and A.~Sagnotti, {\it {Toroidal compactification and
  symmetry breaking in open string theories}},  {\em Nucl.Phys.} {\bf B376}
  (1992) 365--386.

\bibitem{Angelantonj:1996mw}
C.~Angelantonj, M.~Bianchi, G.~Pradisi, A.~Sagnotti, and Ya.~S. Stanev, {\it
  {Comments on Gepner models and type I vacua in string theory}},  {\em
  Phys.Lett.} {\bf B387} (1996) 743--749,
  [\href{http://xxx.lanl.gov/abs/hep-th/9607229}{{\tt hep-th/9607229}}].

\bibitem{Angelantonj:1996uy}
C.~Angelantonj, M.~Bianchi, G.~Pradisi, A.~Sagnotti, and Ya.~Stanev, {\it
  {Chiral asymmetry in four-dimensional open string vacua}},  {\em Phys.Lett.}
  {\bf B385} (1996) 96--102,
  [\href{http://xxx.lanl.gov/abs/hep-th/9606169}{{\tt hep-th/9606169}}].

\bibitem{Pradisi:1995qy}
G.~Pradisi, A.~Sagnotti, and Ya.~S. Stanev, {\it {Planar duality in SU(2) WZW
  models}},  {\em Phys.Lett.} {\bf B354} (1995) 279--286,
  [\href{http://xxx.lanl.gov/abs/hep-th/9503207}{{\tt hep-th/9503207}}].

\bibitem{Recknagel:1997sb}
A.~Recknagel and V.~Schomerus, {\it {D-branes in Gepner models}},  {\em
  Nucl.Phys.} {\bf B531} (1998) 185--225,
  [\href{http://xxx.lanl.gov/abs/hep-th/9712186}{{\tt hep-th/9712186}}].

\bibitem{Blumenhagen:2003su}
R.~Blumenhagen, {\it {Supersymmetric orientifolds of Gepner models}},  {\em
  JHEP} {\bf 0311} (2003) 055,
  [\href{http://xxx.lanl.gov/abs/hep-th/0310244}{{\tt hep-th/0310244}}].

\bibitem{Blumenhagen:2004cg}
R.~Blumenhagen and T.~Weigand, {\it {Chiral supersymmetric Gepner model
  orientifolds}},  {\em JHEP} {\bf 0402} (2004) 041,
  [\href{http://xxx.lanl.gov/abs/hep-th/0401148}{{\tt hep-th/0401148}}].

\bibitem{Dijkstra:2004ym}
T.~Dijkstra, L.~Huiszoon, and A.~Schellekens, {\it {Chiral supersymmetric
  standard model spectra from orientifolds of Gepner models}},  {\em
  Phys.Lett.} {\bf B609} (2005) 408--417,
  [\href{http://xxx.lanl.gov/abs/hep-th/0403196}{{\tt hep-th/0403196}}].

\bibitem{Anastasopoulos:2006da}
P.~Anastasopoulos, T.~Dijkstra, E.~Kiritsis, and A.~Schellekens, {\it
  {Orientifolds, hypercharge embeddings and the Standard Model}},  {\em
  Nucl.Phys.} {\bf B759} (2006) 83--146,
  [\href{http://xxx.lanl.gov/abs/hep-th/0605226}{{\tt hep-th/0605226}}].

\bibitem{Gava:1997jt}
E.~Gava, K.~S. Narain, and M.~H. Sarmadi, {\it {On the bound states of p- and
  (p+2)-branes}},  {\em Nucl. Phys.} {\bf B504} (1997) 214--238,
  [\href{http://xxx.lanl.gov/abs/hep-th/9704006}{{\tt hep-th/9704006}}].

\bibitem{David:2000um}
J.~R. David, {\it {Tachyon condensation in the D0/D4 system}},  {\em JHEP} {\bf
  10} (2000) 004, [\href{http://xxx.lanl.gov/abs/hep-th/0007235}{{\tt
  hep-th/0007235}}].

\bibitem{David:2000yn}
J.~R. David, {\it {Tachyon condensation using the disc partition function}},
  {\em JHEP} {\bf 07} (2001) 009,
  [\href{http://xxx.lanl.gov/abs/hep-th/0012089}{{\tt hep-th/0012089}}].

\bibitem{Cvetic:2003ch}
M.~Cveti{\v c} and I.~Papadimitriou, {\it {Conformal field theory couplings for
  intersecting D-branes on orientifolds}},  {\em Phys. Rev.} {\bf D68} (2003)
  046001, [\href{http://xxx.lanl.gov/abs/hep-th/0303083}{{\tt
  hep-th/0303083}}].

\bibitem{Abel:2003vv}
S.~A. Abel and A.~W. Owen, {\it {Interactions in intersecting brane models}},
  {\em Nucl. Phys.} {\bf B663} (2003) 197--214,
  [\href{http://xxx.lanl.gov/abs/hep-th/0303124}{{\tt hep-th/0303124}}].

\bibitem{Abel:2003yx}
S.~A. Abel and A.~W. Owen, {\it {N-point amplitudes in intersecting brane
  models}},  {\em Nucl. Phys.} {\bf B682} (2004) 183--216,
  [\href{http://xxx.lanl.gov/abs/hep-th/0310257}{{\tt hep-th/0310257}}].

\bibitem{Light}
P.~Anastasopoulos, M.~Bianchi, and R.~Richter, {\it {Light stringy states}}, .

\bibitem{Dixon:1986qv}
L.~J. Dixon, D.~Friedan, E.~J. Martinec, and S.~H. Shenker, {\it {The Conformal
  Field Theory of Orbifolds}},  {\em Nucl. Phys.} {\bf B282} (1987) 13--73.

\bibitem{Burwick:1990tu}
T.~T. Burwick, R.~K. Kaiser, and H.~F. M{\"u}ller, {\it {General Yukawa
  couplings of strings on Z(N) orbifolds}},  {\em Nucl. Phys.} {\bf B355}
  (1991) 689--711.

\bibitem{BianchiPHD}
M.~Bianchi, {\it {PhD thesis}}, .

\bibitem{StanevMaster}
Ya.~Stanev, {\it {Master thesis}}, .

\bibitem{Fuchs:1999fh}
J.~Fuchs and C.~Schweigert, {\it {Bundles of chiral blocks and boundary
  conditions in CFT}},  \href{http://xxx.lanl.gov/abs/hep-th/0001005}{{\tt
  hep-th/0001005}}. Slightly extended version of contribution to the
  Proceedings.

\bibitem{Fuchs:2000hn}
J.~Fuchs and C.~Schweigert, {\it {D-Brane conformal field theory and bundles of
  conformal blocks}},  \href{http://xxx.lanl.gov/abs/math/0004034}{{\tt
  math/0004034}}.

\bibitem{Antoniadis:1993jp}
I.~Antoniadis and K.~Benakli, {\it {Limits on extra dimensions in orbifold
  compactifications of superstrings}},  {\em Phys.Lett.} {\bf B326} (1994)
  69--78, [\href{http://xxx.lanl.gov/abs/hep-th/9310151}{{\tt
  hep-th/9310151}}].

\bibitem{Kors:2001ku}
B.~K{\"o}rs, {\it {Open strings in magnetic background fields}},  {\em Fortsch.
  Phys.} {\bf 49} (2001) 759--867.

\bibitem{Angelantonj:2005hs}
C.~Angelantonj, M.~Cardella, and N.~Irges, {\it {Scherk-Schwarz breaking and
  intersecting branes}},  {\em Nucl. Phys.} {\bf B725} (2005) 115--154,
  [\href{http://xxx.lanl.gov/abs/hep-th/0503179}{{\tt hep-th/0503179}}].

\bibitem{Cremades:2003qj}
D.~Cremades, L.~E. Ib{\'a}{\~n}ez, and F.~Marchesano, {\it {Yukawa couplings in
  intersecting D-brane models}},  {\em JHEP} {\bf 07} (2003) 038,
  [\href{http://xxx.lanl.gov/abs/hep-th/0302105}{{\tt hep-th/0302105}}].

\bibitem{Cremades:2004wa}
D.~Cremades, L.~E. Ib{\'a}{\~n}ez, and F.~Marchesano, {\it {Computing Yukawa
  couplings from magnetized extra dimensions}},  {\em JHEP} {\bf 05} (2004)
  079, [\href{http://xxx.lanl.gov/abs/hep-th/0404229}{{\tt hep-th/0404229}}].

\bibitem{Cvetic:2006iz}
M.~Cveti{\v c} and R.~Richter, {\it {Proton decay via dimension-six operators
  in intersecting D6-brane models}},  {\em Nucl. Phys.} {\bf B762} (2007)
  112--147, [\href{http://xxx.lanl.gov/abs/hep-th/0606001}{{\tt
  hep-th/0606001}}].

\bibitem{Bertolini:2005qh}
M.~Bertolini, M.~Bill{\`o}, A.~Lerda, J.~F. Morales, and R.~Russo, {\it {Brane
  world effective actions for D-branes with fluxes}},  {\em Nucl. Phys.} {\bf
  B743} (2006) 1--40, [\href{http://xxx.lanl.gov/abs/hep-th/0512067}{{\tt
  hep-th/0512067}}].

\bibitem{Cvetic:2009mt}
M.~Cveti{\v c}, I.~Garcia-Etxebarria, and R.~Richter, {\it {Branes and
  instantons intersecting at angles}},  {\em JHEP} {\bf 1001} (2010) 005,
  [\href{http://xxx.lanl.gov/abs/0905.1694}{{\tt arXiv:0905.1694}}].

\bibitem{Bianchi:2010es}
M.~Bianchi, L.~Lopez, and R.~Richter, {\it {On stable higher spin states in
  Heterotic String Theories}},  {\em JHEP} {\bf 1103} (2011) 051,
  [\href{http://xxx.lanl.gov/abs/1010.1177}{{\tt arXiv:1010.1177}}].

\bibitem{Stieberger:1992bj}
S.~Stieberger, D.~Jungnickel, J.~Lauer, and M.~Spalinski, {\it {Yukawa
  couplings for bosonic Z(N) orbifolds: Their moduli and twisted sector
  dependence}},  {\em Mod. Phys. Lett.} {\bf A7} (1992) 3059--3070,
  [\href{http://xxx.lanl.gov/abs/hep-th/9204037}{{\tt hep-th/9204037}}].

\bibitem{Erler:1992gt}
J.~Erler, D.~Jungnickel, M.~Spalinski, and S.~Stieberger, {\it {Higher twisted
  sector couplings of Z(N) orbifolds}},  {\em Nucl. Phys.} {\bf B397} (1993)
  379--416, [\href{http://xxx.lanl.gov/abs/hep-th/9207049}{{\tt
  hep-th/9207049}}].

\end{thebibliography}

\providecommand{\href}[2]{#2}\begingroup\raggedright

\end{document}